\documentclass[11pt]{eptcs}

\usepackage{graphicx}
\usepackage{latexsym}
\usepackage{amsmath}
\usepackage{breakurl}
\usepackage[dvips]{color}

\def\mvn{\emph{MV}}
\def\mvni#1{\mvn_{#1}}
\def\mvnEx1{Ex1}
\def\mvnExB{Ex2}
\def\mvnExC{Ex3}
\def\mvnAI{AB_{1}}
\def\mvnAII{AB_{2}}
\def\nodeAll{ G } 
\def\node#1{ g_{#1} } 
\def\nextSt#1{ [#1] }
\def\modelSS#1{ S_{\scriptstyle #1} }
\def\ss{Y}

\def\ssEntity#1{ \ss(#1) } 

\def\nbh{N}
\def\nbhEntity#1{\nbh(#1)} 
\def\func{F}
\def\funcEntity#1{ f_{ #1 } } 
\def\trace{\sigma}
\def\traceState#1{\trace(#1)}
\def\language#1{Tr(#1)}
\def\absMap{\phi}
\def\mapping#1{\absMap(#1)}

\def\mappingSet#1{MS(#1)}
\def\refines#1{\lhd^{#1}}
\def\exrefines#1{=^{#1}}
\def\setAbs#1#2{\emph{AS}(#1,#2)}
\def\oneStep#1{ \stackrel{{\tiny #1}}{\rightarrow}}
\def\updateStep{\rightarrow}
\def\corrState#1{\lhd^{#1}}
\def\corrTrace#1{\lhd^{#1}}
\def\attor#1{\left\langle #1 \right\rangle}
\def\attorTr#1{\emph{att}(#1)}
\def\ATTOR#1{\emph{ATT}(#1)}
\def\PLI{\emph{PL2}}
\def\PLIab{\emph{APL2}}
\def\PLII{\emph{PL4}}
\def\PLIIabI{\emph{APL4}_{1}}
\def\PLIIabII{\emph{APL4}_{2}}

\def\pfbox{ \hspace*{\fill} $\Box$ }

\newcommand{\CI}     {\mathit{CI}}
\newcommand{\CII}    {\mathit{CII}}
\newcommand{\N}     {\mathit{N}}
\newcommand{\Cro}    {\mathit{Cro}}

\newcounter{defsThms}
\def\addDefn#1#2{{\refstepcounter{defsThms}\label{#2}{\bf #1  \arabic{defsThms}.~}}}

\title{\bf An Abstraction Theory for Qualitative Models \\ of Biological Systems }
\author{Richard Banks
\institute{School of Computing Science, University of Newcastle.}
\email{richard.banks@ncl.ac.uk}
\and
L. Jason Steggles
\institute{School of Computing Science, University of Newcastle.}
\email{l.j.steggles@ncl.ac.uk}
}
\def\titlerunning{An Abstraction Theory for Qualitative Models of Biological Systems}
\def\authorrunning{R. Banks \& L. J. Steggles}

\begin{document}

\maketitle

\begin{abstract}
Multi-valued network models are an important qualitative modelling approach used widely by the biological community.
In this paper we consider developing an abstraction theory for multi-valued network models that
allows the state space of a model to be reduced while preserving key properties of the model.
This is important as it aids the analysis and comparison of multi-valued networks
and in particular, helps address the well--known problem of state space explosion associated with such analysis.
We also consider developing techniques for efficiently identifying abstractions and so provide a basis for the automation of this task.
We illustrate the theory and techniques developed by investigating the identification of abstractions for two
published MVN models of the lysis--lysogeny switch in the bacteriophage $\lambda$.
\end{abstract}

\section{Introduction}
In order to understand and analyse the complex control mechanisms inherent in biological systems
a range of formal modelling techniques have been applied by biologists (for an overview see for example \cite{Bower01,deJong02}).
In particular, qualitative modelling techniques have emerged as an important modelling approach
due to the lack of quantitative data on reaction rates and the noise associated with such data.
\emph{Multi-valued networks} (MVNs) \cite{Rudell1987,Thomas1990,Thomas1995} are a promising qualitative modelling approach
for biological systems.
They extend the well--known \emph{Boolean network} approach \cite{amk99,Bower01} by allowing the state of each regulatory entity to be
within a range of discrete values instead of just on/off.
In this way they are able to provide a compromise between the simplicity of Boolean networks and
the more detailed differential equational models.

However, the analysis of MVNs is not without problems.
They suffer from the well--known state space explosion problem, a problem which is exacerbated in MVNs
by the possibly large set of states associated with each individual entity.
Another important shortcoming is the lack of any techniques for relating MVN models at different levels of abstraction.
This hinders the comparison of MVN models and means there is no basis for the incremental development of MVNs.

In this paper we begin to address these problems by developing an abstraction theory for MVNs.
Abstraction techniques are a well established approach in the area of formal verification
(see for example \cite{Clarke1994,Bensalem1998}) which allow a simpler model to be identified
which can then be used to provide insight into the more complex original model.
The abstraction theory we present is based on using an abstraction mapping to relate the reduced state space of
an abstraction to the original MVN model.
We develop a notion of what it means for one MVN to correctly abstract another and investigate the
scope and limits of the analysis properties that can be inferred from an abstraction model.
We show that abstractions allow sound analysis inferences about reachability properties
in the sense that any reachability result shown on the abstraction must hold on the original model.
Importantly, we show that all attractors of an abstraction correspond to attractors in the original model.


We illustrate the theory and techniques developed by
investigating the existence of abstractions for two published MVN models
for the genetic regulatory network controlling the lysis--lysogeny
switch in the bacteriophage $\lambda$ \cite{Thieffry1995, Chaouiya2008}.
Bacteriophage $\lambda$ \cite{Thomas1990, opp05} is a
virus which after infecting the bacteria \emph{Escherichia coli}
makes a decision to switch to one of two possible reproductive phases.
It can enter the \emph{lytic cycle} where the virus generates as many new viral particles as the
infected cell's resources allow and then lyse
the cell wall to release the new phage.
Alternatively, it can enter the \emph{lysogenic cycle} where the $\lambda$ DNA integrates into the host
DNA providing it with immunity from other phages and allowing it to be replicated with each cell division.
We consider a two and four entity MVN model \cite{Thieffry1995} of the lysis--lysogeny switching mechanism
and using the techniques we have developed identify corresponding abstractions for these models.

This paper is organized as follows.
In Section 2 we provide a brief overview of the MVN modelling approach and
present a simple illustrative example.
In Section 3 we develop an abstraction theory for multi-valued networks and
present a range of results concerning this theory.
In Section 4 we consider the identification of abstractions and develop a basis
for automating the abstraction process.
In Section 5 we illustrate the theory and techniques developed by presenting
two abstraction examples for published models of the lysis--lysogeny switch
in bacteriophage $\lambda$.
Finally, in Section 6 we present some concluding remarks and consider directions for future work. 

\section{Multi-valued Network Models}
\label{sec:mvn}
In this section, we introduce {\it multi-valued networks} (MVNs) \cite{Rudell1987,Thomas1990,Thomas1995},
a qualitative modelling approach which extends the well-known \emph{Boolean network} \cite{amk99} approach
by allowing the state of each regulatory entity to be within a range of discrete values.
MVNs have been extensively studied in circuit design (for example, see \cite{Rudell1987,Mishchenko2002}) and successfully applied to modelling biological systems (for example, see \cite{Thomas1995,Chaouiya2008,Schaub2007}).

An MVN consists of a set of logically linked entities $G = \{\node{1} , \ldots, \node{k} \}$ which regulate each other in a positive or negative way.
Each entity $\node{i}$ in an MVN has an associated set of discrete states $\ssEntity{\node{i}} = \{0,\dots, m_{i}\}$, for some $m_{i} \geq 1$, from which its current state is taken.
Note that a Boolean network is therefore simply an MVN in which each entity $\node{i}$ has a Boolean set of states $\ssEntity{\node{i}} = \{0, 1\}$.
Each entity $\node{i}$ also has a neighbourhood $\nbhEntity{\node{i}} = \{\node{i_{1}}, \ldots, \node{i_{l(i)}} \}$
which is the set of all entities that can directly affect its state. (Note that $\node{i}$ may or may not
be a member of $\nbhEntity{\node{i}}$.)
Furthermore, interactions between one entity and another only become functional if the state of the source entity
has reached some threshold state level (this threshold state level is always at least one).
MVNs can therefore discriminate between the strengths of different interactions, something which Boolean networks are unable to capture.
The behaviour of each entity $\node{i}$ based on these neighbourhood interactions is formally defined by a logical next-state function
$\funcEntity{\node{i}}$ which calculates the next-state of $\node{i}$ given the current states of the entities in its neighbourhood.

We can now define an MVN more formally as follows.
\\
\\
\addDefn{Definition}{def:mvn}
An MVN $\mvn$ is a four-tuple $\mvn = (\nodeAll, \ss, \nbh, \func)$ where:
\\
i) $\nodeAll=\{ \node{1} , \dots, \node{k} \}$ is a non-empty, finite set of entities;
\\
ii) $\ss = \left( \ssEntity{ \node{1}} , \ldots, \ssEntity{\node{k}} \right)$ is a tuple of state sets,
where each $\ssEntity{\node{i}} = \{0, \ldots, m_{i}\}$, for some $m_{i} \geq 1$, is the state space for entity $\node{i}$;
\\
iii) $\nbh = \left( \nbhEntity{ \node{1}}  , \ldots, \nbhEntity{ \node{k}} \right) $ is a tuple of neighbourhoods,
such that $\nbhEntity{ \node{i}}  \subseteq \nodeAll $ is the neighbourhood of $\node{i}$; and
\\
iv) $\func = \left( \funcEntity{\node{1}}, \dots, \funcEntity{\node{k}} \right) $ is a tuple of next-state multi-valued functions, such that if
$\nbhEntity{\node{i}} = \{\node{i_1} ,\ldots, \node{i_n} \}$ then the function
$\funcEntity{\node{i}} : \ssEntity{\node{i_1}} \times \dots \times \ssEntity{\node{i_n}} \rightarrow \ssEntity{\node{i}}$
defines the next state of $\node{i}$.
\pfbox
\\

In the sequel, let $\mvn = (\nodeAll, \ss, \nbh, \func)$ be an arbitrary MVN.
In a slight abuse of notation we let $\node{i} \in \mvn$ represent that $\node{i} \in \nodeAll$ is an entity in $\mvn$.

\begin{figure}[h]
\centering
\begin{tabular}{c@{\qquad}c}
  \begin{tabular}{c}

   \includegraphics[width=0.3\textwidth,keepaspectratio]{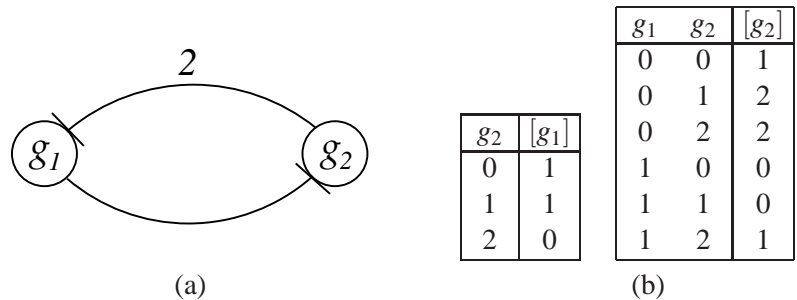}
  \end{tabular}
        &
    \begin{tabular}{c@{\qquad}c@{\qquad}c}
\begin{tabular}[b]{|c|c|}
        \hline $\node{2} $  & $\!\!\nextSt{\node{1} }\!\!$  \\
        \hline
        0 &  1\\
        1 &  1 \\
        2 &  0 \\
        \hline
\end{tabular}
    & \hspace{-1.0em}
\begin{tabular}[b]{|cc|c|}
        \hline $\node{1} $  & $\node{2} $ & $\!\!\nextSt{\node{2} }\!\!$\\
        \hline
        0 & 0 & 1 \\
        0 & 1 & 2 \\
        0 & 2 & 2 \\
        1 & 0 & 0 \\
        1 & 1 & 0 \\
        1 & 2 & 1 \\
        \hline
    \end{tabular}

\end{tabular}
    \\[-0.1em]
    (a) & (b) \\
\end{tabular}
\caption{
An example MVN $\mvnEx1$ which consists of two entities $\node{1}$ and $\node{2}$, including:
(a) network structure; and (b) the state transition tables representing the corresponding next-state functions.}
\label{fig:mvnExample}
\end{figure}

As an example, consider the MVN $\mvnEx1$ defined in Figure \ref{fig:mvnExample}
which consists of two entities $\node{1}$ and $\node{2}$,
such that $\ssEntity{\node{1}} = \{0,1\}$ and $\ssEntity{\node{2}} = \{0,1,2\}$.
The update functions for each entity are defined using state transition tables (see Figure \ref{fig:mvnExample}.(b)) where $\nextSt{\node{i}}$ is used to denote
the next state of an entity $\node{i}$.
It can be seen that entity $\node{1}$ inhibits $\node{2} $ and
that entity $\node{2}$ inhibits  $\node{1}$ but only when it reaches state $2$
(this is represented in Figure \ref{fig:mvnExample}.(a) by labelling the corresponding edge with a 2).
Note that although $\node{2} \in \nbhEntity{\node{2}}$ we have not drawn an edge for this in Figure \ref{fig:mvnExample}.(a) since
$\node{2}$ has no regulatory affect on itself and is needed simply to allow the affect of $\node{1}$ to be precisely defined.

A \emph{global state} of an MVN $\mvn$ with $k$ entities is represented
by a tuple of states $(s_{1}, \ldots, s_{k})$,
where $s_{i} \in \ssEntity{\node{i}}$ represents the state of entity $\node{i} \in \mvn$.
Note as a notational convenience we often use $s_{1} \ldots s_{k}$
to represent a global state $(s_{1}, \ldots, s_{k})$.
When the current state of an MVN is clear from the context we allow $\node{i}$ to denote both the name of an entity and its corresponding current state.
The state space of an MVN $\mvn$, denoted $\modelSS{\mvn}$, is therefore the set of all possible global states
$\modelSS{\mvn}  = \ssEntity{\node{1}} \times \cdots \times \ssEntity{\node{k}}$.
The state of an MVN can be updated either \emph{synchronously}, where the state of all entities is updated simultaneously in a single update step, or
\emph{asynchronously}, where entities update their state independently (see \cite{Harvey1997}).
In the following we focus on the synchronous update semantics since this has received considerable attention from the biological community.
Given two states $S_{1}, S_{2} \in \modelSS{\mvn}$, let $S_{1} \updateStep S_{2}$ represent a \emph{synchronous update step}
such that $S_{2}$ is the state that results from simultaneously updating the state
of each entity $\node{i}$ using its associated update function $f_{\node{i}}$ and the appropriate neighbourhood of states from $S_{1}$.

As an example, consider the global state $01$ for $\mvnEx1$ (see Figure \ref{fig:mvnExample}) in which $g_{1}$ has state $0$
and $g_{2}$ has state $1$.
Then
$01 \updateStep 12$
is a single synchronous update step on this state resulting in the new state $12$.
%
%
The sequence of global states through $\modelSS{\mvn} $ from some initial state is called a \emph{trace}.
Note that in the case of a synchronous update semantics such traces are infinite.
However, given that the global state space is finite, this implies that a trace must eventually enter a cycle,
known formally as an \emph{attractor cycle} \cite{kauffman1993,Thomas1995}.
We make use of this fact to define a finite canonical representation for traces
which specifies a trace up to the first repeated state.
\\
\\
\addDefn{Definition}{def:trace}
Let $S_{0} \in \modelSS{\mvn}$ be a global state for $\mvn$.
A \emph{trace} is a list of global states
$\traceState{S_{0}} = \left\langle S_{0}, S_{1}, \dots, S_{n} \right\rangle$
such that:
\\
i) $S_{i} \updateStep S_{i+1}$, for $0 \leq i < n$;
\\
ii) $S_{0},\dots,S_{n-1}$ are unique states; and
\\
iii) $S_{n}=S_{i}$ for some $i \in \{0,\dots,n-1\}$.
\pfbox
\\

The set of all traces
$\language{\mvn} = \{\traceState{S} \ |\ S \in \modelSS{\mvn} \}$
therefore completely characterizes the behaviour of an MVN model
under the synchronous semantics and is referred to as the \emph{trace semantics} of $\mvn$.

\begin{figure}[h]
\centering
\begin{tabular}[b]{c@{\qquad}c}
\begin{tabular}[b]{l l}
$\traceState{00} = \left\langle 00, 11, 10, 10  \right\rangle$ &
$\traceState{10} = \left\langle 10, 10  \right\rangle$ \\
$\traceState{01} = \left\langle 01, 12, 01  \right\rangle$ &
$\traceState{11} = \left\langle 11, 10, 10  \right\rangle$ \\
$\traceState{02} = \left\langle 02, 02  \right\rangle$ &
$\traceState{12} = \left\langle 12, 01, 12  \right\rangle$ \\
\\
\end{tabular}

  &
\includegraphics[width=0.25\textwidth,keepaspectratio]{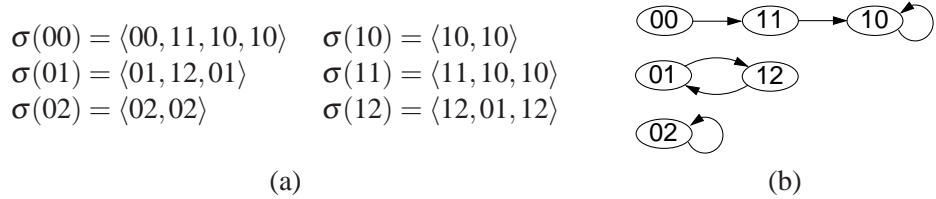}

    \\
    (a) & (b) \\
\end{tabular}
\caption{
The trace semantics for $\mvnEx1$: (a) the set of formal traces; and (b) a graphical representation of the traces.}
\label{fig:traceExample}
\end{figure}

In our running example, $\mvnEx1$ has a state space of size $|\modelSS{\mvnEx1} | = 6$ and
so (under a synchronous update semantics) $\language{\mvnEx1}$ consists of the
six traces presented in Figure \ref{fig:traceExample}.(a) below.

As mentioned above, each trace leads to a cyclic sequence of states known as an \emph{attractor cycle} \cite{kauffman1993,Thomas1995}.
For example, in Figure \ref{fig:traceExample}.(b) we can see that $\mvnEx1$ has
three attractors: $10 \updateStep 10$ and $02 \updateStep 02$ known as \emph{point attractors}; and $01 \updateStep 02 \updateStep 01$
which is an attractor cycle of period 2 \cite{kauffman1993}.

Given a trace $\sigma = \left\langle S_{1}, \ldots, S_{n} \right\rangle \in \language{\mvn}$ for an MVN $\mvn$
we let $\attorTr{\sigma}$ denote the attractor cycle that must occur in trace $\sigma$, i.e.
$\attorTr{\sigma} = \attor{ S_{k}, \ S_{k+1}, \ldots, S_{n} }$,
for some $1 \leq k < n$ and $S_{k} = S_{n}$.
We let $\ATTOR{\mvn}$ denote the set of all attractors for $\mvn$, i.e.
$$
\ATTOR{\mvn} = \{ \attorTr{\sigma} \ | \ \sigma \in \language{\mvn} \}.
$$

Attractor cycles are very important biologically where they are seen as representing different biological states or functions
(e.g. different cellular types such as proliferation, apoptosis and differentiation \cite{huang2000}).
Thus, the identification and analysis of attractor cycles for MVNs is an important subject which has
warranted much attention in the literature (for example, see \cite{kauffman1993,Thomas1995,Drossel2005}).

\section{An Abstraction Theory for MVNs}
\label{sec:refine}
In this section we develop a notion of abstraction for MVNs by considering what it means for
one MVN to abstractly implement the behaviour of another.
This is based around the idea of showing that the trace semantics of one MVN is consistent with the trace semantics of a more
complex MVN under an appropriate mapping of states.


We begin by defining how an entity's state space can be simplified using a mapping to merge states.
\\
\\
\addDefn{Definition}{def:sm}
Let $\mvn$ be an MVN and
let $\node{i} \in \mvn$ be an entity such that $\ss(\node{i}) = \{0, \ldots, m \}$ for some $m > 1$.
Then a \emph{state mapping}
$\mapping{\node{i}}$ for entity $\node{i}$
is a surjective mapping
$\mapping{\node{i}} : \{0, \ldots, m \} \rightarrow\ \{0, \ldots,n \}$, where $0 < n < m$.
\pfbox
\\

The idea is that a state mapping reduces the set of states an entity can be in by merging appropriate states.
The state mapping must be surjective to ensure that all states in the new reduced state space are used.
Note we only consider state mappings with a codomain larger than one, since a singular state
entity does not appear to be of biological interest.

As an example, consider entity $\node{2} \in \mvnEx1$ (see Figure \ref{fig:mvnExample}) which has the state space
$\ssEntity{\node{2}} = \{0, 1, 2 \}$.
It is only meaningful to simplify $\node{2} \in \mvnEx1$ to a Boolean entity and so
one possible state mapping to achieve this would be:
$$
\mapping{\node{2}} = \{0 \mapsto 0, 1 \mapsto 0, 2 \mapsto 1\},
$$
which merges states $0$ and $1$ into a single state $0$, and translates state $2$ into $1$.

Clearly, there are a number of different possible state mappings which can be applied to reduce a node's state space
from $m$ to $n$ states, for $1 < n < m$.
The complete set of all such state mappings is denoted
$\mappingSet{m,n} = \{ \ \phi \ | \ \phi : \{0, \ldots, m-1 \} \rightarrow\ \{0, \ldots, n-1 \} \ and \ \phi \ is \ surjective \}$.
For example, the mapping set $\mappingSet{3,2}$ consists of the following six mappings:
\begin{center}
\begin{tabular}[b]{l l}
  (1)\ \ $\{0 \mapsto 0, 1 \mapsto 0, 2 \mapsto 1\}$ \ \ \ \ \ &
  (4)\ \ $\{0 \mapsto 1, 1 \mapsto 1, 2 \mapsto 0\}$ \\
  (2)\ \ $\{0 \mapsto 0, 1 \mapsto 1, 2 \mapsto 1\}$  &
  (5)\ \ $\{0 \mapsto 1, 1 \mapsto 0, 2 \mapsto 0\}$ \\
  (3)\ \ $\{0 \mapsto 0, 1 \mapsto 1, 2 \mapsto 0\}$ &
  (6)\ \ $\{0 \mapsto 1, 1 \mapsto 0, 2 \mapsto 1\}$
\end{tabular}
\end{center}

In order to be able to consider simplifying several entities at the same time during the abstraction process
we introduce the notion of a family of state mappings as follows.
\\
\\
\addDefn{Definition}{def:absMap}
Let $\mvn = (\nodeAll, \ss, \nbh, \func)$ be an MVN with entities $G = \{ \node{1} , \dots, \node{k} \}$.
Then an \emph{abstraction mapping}
$\absMap$ for $\mvn$
is a family of mappings
$\absMap = \langle \mapping{\node{1}}, \ldots, \mapping{\node{k}} \rangle$
such that for each $1 \leq i \leq k$ we have
$\mapping{\node{i}}$ is either a state mapping for entity $\node{i}$ or is the identity mapping
$I_{\node{i}} : \ss(\node{i}) \rightarrow \ss(\node{i})$
where $I_{\node{i}}(s) = s$,
for all $s \in \ss(\node{i})$.
Furthermore, for $\absMap$ to be well--defined we insist that at least one of the mappings
$\mapping{\node{i}}$ is a state mapping.
\pfbox
\\

Note in the sequel given a state mapping $\mapping{\node{i}}$ we let it denote both itself and the
corresponding abstraction mapping containing
only the single state mapping $\mapping{\node{i}}$.

An abstraction mapping can be lifted and applied to the trace semantics of an MVN as follows.
\\
\\
\addDefn{Definition}{dfn:mappingApp}
An abstraction mapping $\absMap = \langle \mapping{\node{1}} \ldots \mapping{\node{k}} \rangle$ for $\mvn$
can be used to abstract a global state
$s_{1} \ldots s_{k} \in \modelSS{\mvn}$
by applying it pointwise, i.e.
$\absMap(s_{1}\dots s_{k}) = \mapping{\node{1}}(s_{1}) \ldots \mapping{\node{k}} (s_{k})$.
We can lift an abstraction mapping $\absMap$ to a trace
$\traceState{S_{0}} = \left \langle S_{0}, \ldots, S_{n} \right \rangle \in \language{\mvn}$
by applying $\absMap$ to each global state in the trace as follows
$$
\absMap(\traceState{S_{0}}) =
\left\langle \absMap(S_{0}), \ldots, \absMap(S_{n}) \right\rangle.
$$
However, $\absMap(\traceState{S_{0}})$ may contain contradictory steps and thus not represent a meaningful abstracted trace.
We say an abstracted trace $\absMap(\traceState{S_{0}})$ is \emph{valid} iff there does not exist two identical states $\absMap(S_{i}) = \absMap(S_{j})$,
for some $i,j \in \{ 0,\ldots,n-1 \}$, such that
$\absMap(S_{i+1}) \neq \absMap(S_{j+1})$.
If $\absMap(\traceState{S_{0}})$ is a valid abstracted trace then we need to ensure it is in the canonical form introduced in Definition \ref{def:trace}.
We do this by removing any repeating tail that may have been introduced by the abstraction mapping, i.e.
choose the smallest $k$, $0 < k \leq n$ such that
$\absMap(S_{0}), \ldots, \absMap(S_{k-1})$ are unique states
and $\absMap(S_{i}) = \absMap(S_{k})$, for some $i \in \{ 0,\ldots,k-1 \}$.
(Note whenever we talk about a valid abstracted trace we will assume it is in its canonical form.)

We can lift $\absMap$ to the trace semantics of a model $\mvn$:
$$
\absMap(\language{\mvn}) = \{\absMap(\traceState{S}) \ | \ \traceState{S} \in \language{\mvn} \text{ and } \absMap(\traceState{S}) \text{ is valid}  \}.
$$
\pfbox
\\

Continuing with our running example, $\mapping{\node{2}}$ can be applied as an abstraction mapping to the
trace semantics $\language{\mvnEx1}$
(see Figure \ref{fig:traceExample}) resulting in the abstracted trace semantics $\mapping{\node{2}}(\language{\mvnEx1})$,
shown below in Figure \ref{fig:abstractTrace},
in which the states of $\node{2}$ have been reduced accordingly.
\begin{figure}[h]
\label{fig:abstractTrace}
\begin{center}
\begin{tabular}[b]{l c l}
$\mapping{\node{2}} (\traceState{00}) = \left\langle 00, 10, 10  \right\rangle$ & &
$\mapping{\node{2}} (\traceState{10}) = \left\langle 10, 10  \right\rangle$ \\
$\mapping{\node{2}} (\traceState{01}) = \left\langle 00, 11, 00  \right\rangle$ & &
$\mapping{\node{2}} (\traceState{11}) = \left\langle 10, 10  \right\rangle$ \\
$\mapping{\node{2}} (\traceState{02}) = \left\langle 01, 01  \right\rangle$ & &
$\mapping{\node{2}} (\traceState{12}) = \left\langle 11, 00, 11  \right\rangle$
\end{tabular}
\end{center}
\caption{The trace semantics $\mapping{\node{2}}(\language{\mvnEx1})$ resulting from abstracting $\language{\mvnEx1}$ using $\mapping{\node{2}}$. }
\end{figure}

Note that $\mapping{\node{2}} (\language{\mvnEx1})$
is non--deterministic in the sense that we have two different traces beginning with the same state $00$
(i.e. starting in state $00$ we have a non-deterministic choice between two abstracted traces,
$\left\langle 00, 10, 10  \right\rangle$ and $\left\langle 00, 11, 00  \right\rangle$).
This occurs as we are viewing the more complex set of behaviours captured by $\language{\mvnEx1}$ from a simpler perspective.

To illustrate how invalid abstracted traces arise consider an MVN with two entities that has the following trace
$\traceState{00} = \left\langle 00, 11, 01, 02, 02 \right\rangle$.
When $\traceState{00}$ is abstracted with the standard abstraction mapping
$\mapping{\node{2}} =\{0 \mapsto 0, 1 \mapsto 0, 2 \mapsto 1 \}$
the result is the following
$$
\mapping{\node{2}}(\traceState{00}) = \left\langle 00, 10, 00, 01, 01 \right\rangle.
$$
However, it can be observed that this is not a valid trace according to Definition \ref{dfn:mappingApp}
because global state $00$ can lead to two different states and will therefore be omitted from
the abstracted trace semantics.

We are now ready to define what it means for one MVN to be an abstraction of another.
\\
\\
\addDefn{Definition}{def:refinement}
Let $\mvni{1} = (\nodeAll_{1}, \ss_{1}, \nbh_{1}, \func_{1})$ and
$\mvni{2} = (\nodeAll_{2}, \ss_{2}, \nbh_{2}, \func_{2})$ be two MVNs with the same structure,
i.e. $\nodeAll_{1} = \nodeAll_{2}$ and $\nbh_{1}(\node{i}) = \nbh_{2}(\node{i})$, for all $\node{i} \in \mvni{1}$.
Let $\absMap$ be an abstraction mapping from $\mvni{2}$ to $\mvni{1}$.
Then we say that
$\mvni{1}$ \emph{abstracts} $\mvni{2}$ \emph{under} $\absMap$,
denoted
$\mvni{1} \refines{\absMap} \mvni{2}$,
if, and only if,
$\language{\mvni{1}} \subseteq \absMap(\language{\mvni{2}})$.
\pfbox
\\

An abstraction
$\mvni{1} \refines{\absMap} \mvni{2}$
indicates that the model $\mvni{1}$ consistently abstracts the behaviour of a more complex model $\mvni{2}$
by reducing the state space of those entities identified in the abstraction mapping $\absMap$.
Note alternatively, we could consider $\mvni{2}$ to be a \emph{refinement} of $\mvni{1}$ in the sense
that $\mvni{2}$ consistently extends $\mvni{1}$ with the addition of further states.
Such a notion of refinement is useful as it provides a framework for the incremental development of MVN models.

As an abstraction example, consider the MVN  $\mvnExB$ defined in Figure \ref{fig:mvnEx2}
which has the same structure as $\mvnEx1$ (see Figure \ref{fig:mvnExample}) but is a Boolean model.
\begin{figure}[h]
\centering
\begin{tabular}{c@{\qquad}c}
  \begin{tabular}{c@{\qquad}c}
     \begin{tabular}[b]{|c|c|}
        \hline $\node{2} $  & $\!\!\nextSt{\node{1} }\!\!$  \\
        \hline
        0 & 1 \\
        1 & 0 \\
        \hline
     \end{tabular}
    & \hspace{-1.0em}
     \begin{tabular}[b]{|cc|c|}
        \hline $\node{1} $  & $\node{2} $ & $\!\!\nextSt{\node{2} }\!\!$\\
        \hline
        0 & 0 & 1 \\
        0 & 1 & 1 \\
        1 & 0 & 0 \\
        1 & 1 & 0 \\
        \hline
    \end{tabular}
  \end{tabular}
    &
  \begin{tabular}{l}
    $\traceState{00} = \left\langle 00, 11, 00  \right\rangle$ \\
    $\traceState{01} = \left\langle 01, 01  \right\rangle$ \\
    $\traceState{10} = \left\langle 10, 10  \right\rangle$ \\
    $\traceState{11} = \left\langle 11, 00, 11  \right\rangle$
  \end{tabular}
\end{tabular}
\caption{State transition tables defining $\mvnExB$ and its associated trace semantics $\language{\mvnExB}$.}
\label{fig:mvnEx2}
\end{figure}
Then clearly, given the abstraction mapping $\mapping{\node{2}}$ introduced earlier, we can see that
$\language{\mvnExB} \subseteq \mapping{\node{2}}(\language{\mvnEx1})$
holds and so $\mvnExB$ is an abstraction of $\mvnEx1$,
i.e. $\mvnExB \refines{\mapping{\node{2}}} \mvnEx1$ holds.

In special cases, an abstraction may exactly capture the behaviour of the original MVN model under the given abstraction mapping.
We distinguish this stronger case with the notion of an \emph{exact abstraction}.
\\
\\
\addDefn{Definition}{def:exactRefinement}
Let $\mvni{1}$ and
$\mvni{2}$ be two MVNs such that
$\mvni{1} \refines{\absMap} \mvni{2}$
for some abstraction mapping $\absMap$.
Then we say that
$\mvni{1}$ \emph{exactly abstracts} $\mvni{2}$ \emph{under} $\absMap$,
denoted
$\mvni{1} \exrefines{\absMap} \mvni{2}$,
if, and only if,
$\language{\mvni{1}} \ = \ \absMap(\language{\mvni{2}})$
and
for every $\sigma \in \language{\mvni{2}}$, the abstracted trace $\absMap(\sigma)$ is valid.
\pfbox
\\

Exact abstractions are interesting as they indicate redundant states (normally corresponding to entity thresholds) which have
no affect on the qualitative behaviour of an MVN.
Subsequently, an exact abstraction provides a simpler representation of an MVN whilst preserving all
its behaviour under the given abstraction mapping.

It is natural to consider whether every (non--Boolean)\footnote{An MVN is said to be non--Boolean if it contains at least one
entity which has more than two possible states. }
MVN has an abstraction.
In other words, do there exist MVNs which contain regulatory interactions which are too subtle to be represented
in a simpler state domain.
This is an interesting question since it provides insight into the need for non-Boolean MVN models.
Unsurprisingly, it turns out that abstractions do not always exist, as formalized in the following theorem.
\\
\\
\addDefn{Theorem}{theorem:notallmv}
Not every non--Boolean MVN has an abstraction.
\\
\\
\textbf{Proof.}
We simply construct a non--Boolean MVN which we show has no abstractions.
Let $\mvnExC$ be defined by extending $\mvnEx1$ (Figure \ref{fig:mvnExample}) with a third Boolean entity
$\node{3}$ which is inhibited whenever $\node{2}$ is in a state greater than or equal to $1$.
The complete definition for $\mvnExC$ is given in Figure \ref{fig:mvnExample4}.
\begin{figure}[h]
\centering
    \begin{tabular}{c@{\qquad}c@{\qquad}c}
\begin{tabular}[b]{|c|c|}
        \hline $\node{2} $  & $\!\!\nextSt{\node{1} }\!\!$  \\
        \hline
        0 & 1 \\
        1 & 1 \\
        2 & 0 \\
        \hline
\end{tabular}
    & \hspace{-1.0em}
\begin{tabular}[b]{|cc|c|}
        \hline $\node{1} $  & $\node{2} $ & $\!\!\nextSt{\node{2} }\!\!$\\
        \hline
        0 & 0 & 1 \\
        0 & 1 & 2 \\
        0 & 2 & 2 \\
        1 & 0 & 0 \\
        1 & 1 & 0 \\
        1 & 2 & 1 \\
        \hline
    \end{tabular}
    & \hspace{-1.0em}
    \begin{tabular}[b]{|c|c|}
        \hline $\node{2} $  & $\!\!\nextSt{\node{3} }\!\!$  \\
        \hline
        0 & 1 \\
        1 & 0 \\
        2 & 0 \\
        \hline
\end{tabular}
         \vspace*{1.0em}
\end{tabular}
\vspace*{-1.3em}
\caption{
The state transition tables defining $\mvnExC$ (used to prove Theorem \ref{theorem:notallmv}). }
\label{fig:mvnExample4}
\end{figure}
We can see that $\node{2} \in \mvnExC$ now acts in two subtly different ways:
on one hand $\node{1}$ is inhibited when $\node{2} = 2$;
and on the other hand, $\node{3}$ is inhibited when $\node{2} \geq 1$.
We can show that no abstraction exists for this model by exhaustively considering
each possible abstraction mapping $\mapping{\node{2}}$ and showing that for every
possible candidate abstraction model $\mvni{A}$ we have
$\language{\mvni{A}} \not \subseteq \mapping{\node{2}}(\language{\mvnExC})$.
\pfbox
\\

This is an important result which, although centered around the relationship assumption formalized by our abstraction theory,
provides insight into the expressive power of MVNs and
in particular, motivates the need for multi-valued modelling techniques.

One of the main motivations for defining an abstraction theory is to allow simplified models of an MVN
to be identified to aid the analysis process.
This therefore raises the question of what properties of an abstraction are preserved by the original MVN and
we end this section by considering this question.

We begin by introducing a notion of corresponding states and traces.
\\
\\
\addDefn{Definition}{def:analogousState}
Let $\mvn$ be an MVN with an abstraction $\mvni{A}$ under a given abstraction mapping $\absMap$,
i.e. $\mvni{A} \refines{\absMap} \mvn$.
Let $S^{A} \in \modelSS{\mvni{A}}$ be some global state of abstraction $\mvni{A}$ and
$S \in \modelSS{\mvn}$ be a global state of the original model $\mvn$.
Then we say that $S^{A}$ \emph{and} $S$ \emph{correspond with respect to} $\absMap$,
denoted $S^{A} \corrState{\absMap} S$,
if, and only if,
$
S^{A} = \absMap(S).
$
Furthermore, given traces
$\sigma^{A} \in \language{\mvni{A}}$ and
$\sigma \in \language{\mvn}$
we say $\sigma^{A}$ \emph{and} $\sigma$ \emph{correspond with respect to} $\absMap$,
denoted $\sigma^{A} \corrTrace{\absMap} \sigma$,
if, and only if,
$\absMap(\sigma)$ is valid and
$
\sigma^{A} = \absMap(\sigma).
$
\pfbox
\\

Let $S_{1} \oneStep{*} S_{2}$ denote the fact that
global state $S_{2} \in \modelSS{\mvn}$ is reachable from global state $S_{1} \in \modelSS{\mvn}$ in the model $\mvn$.
We now clarify the relationship between reachability properties in an abstraction and its corresponding original MVN model.
\\
\\
\addDefn{Theorem}{theorem:reachability}
Let $\mvni{A} \refines{\absMap} \mvn$ for some mapping abstraction $\absMap$
and let $S_{1}^{A}, S_{2}^{A} \in \modelSS{\mvni{A}}$.
If $S_{1}^{A} \oneStep{*} S_{2}^{A}$ in $\mvni{A}$ then there must exist states
$S_{1}, S_{2} \in \modelSS{\mvn}$ such that
$S_{1}^{A} \corrState{\absMap} S_{1}$,
$S_{2}^{A} \corrState{\absMap} S_{2}$, and
$S_{1} \oneStep{*} S_{2}$ in $\mvn$.
\\
\\
\textbf{Proof.}
Since $S_{1}^{A} \oneStep{*} S_{2}^{A}$ there must exist a trace
$\traceState{S_{1}^{A}} \in \language{\mvni{A}}$ containing $S_{2}^{A}$.
From Definition \ref{def:refinement}, we know that
$\language{\mvni{A}} \subseteq \absMap(\language{\mvn})$ must hold.
Therefore there must exist a state $S_{1} \in \modelSS{\mvn}$ such that
$\traceState{S_{1}^{A}} \corrTrace{\absMap} \traceState{S_{1}}$,
i.e. $\absMap(\traceState{S_{1}}) = \traceState{S_{1}^{A} }$.
From this it is straightforward to see that there must exist the
required state $S_{2}$
in trace $\traceState{S_{1}}$ such that
$S_{2}^{A} \corrState{\absMap} S_{2}$ and
$S_{1} \oneStep{*} S_{2}$.
\pfbox
\\

In other words, reachability properties of abstractions have corresponding reachability properties in the original MVN.
However, since abstractions normally capture less behaviour than the original model,
there are limitation on what can be deduced from an abstraction.
It turns out that determining reachability in a model using an abstraction is a \emph{semi-decidable} property:
(i) By Theorem \ref{theorem:reachability} we know that if one state is reachable from another in an abstraction then a corresponding
reachability property must hold in the original model;
(ii) However, if one state is not reachable from another in an abstraction then a corresponding
reachability property in the original MVN may or may not hold and more analysis will be required.

The final result we present is important as it shows that the
attractor cycles found in an abstraction are preserved by the original MVN.
\\
\\
\addDefn{Theorem}{theorem:attractors}
Let $\mvni{A} \refines{\absMap} \mvn$ for some abstraction mapping $\absMap$.
Then
$$
\ATTOR{\mvni{A}} \subseteq \absMap(\ATTOR{\mvn}).
$$
\textbf{Proof.}
Let $\tau \in \ATTOR{\mvni{A}}$ then we need to show that
$\tau \in \absMap(\ATTOR{\mvn})$.
By definition we know there must exist a trace
$\sigma_{A} \in \language{\mvni{A}}$ such that $\attorTr{\sigma_{A}} = \tau$.
Since $\mvni{A} \refines{\absMap} \mvn$ we know there must exist a trace
$\sigma \in \language{\mvn}$ such that $\absMap(\sigma)$ is valid and
$\sigma_{A} = \absMap(\sigma)$.
It follows that $\tau = \absMap(\attorTr{\sigma})$
and so by definition we know that
$\tau \in \absMap(\ATTOR{\mvn})$
as required.
\pfbox 

\section{Identifying Model Abstractions}
\label{sec:ident}
In the previous section we defined a formal notion of what it means for one MVN to be a correct abstraction
of another.
Given an MVN $\mvn$ and an abstraction mapping $\absMap$
we can therefore define the set $\setAbs{\mvn}{\absMap}$ of all
abstractions of $\mvn$ under $\absMap$, i.e.
$$
\setAbs{\mvn}{\absMap} = \{\mvni{A} \ | \ \mvni{A} \refines{\absMap} \mvn \}.
$$
Finding abstractions, i.e. members of $\setAbs{\mvn}{\absMap}$,
is clearly an important task given that they provide a means of
simplifying the analysis of a model and
can help address the well-known problem of state space explosion.
However, in practice, the brute force derivation of this refinement set becomes intractable for all but the smallest MVN.
Specifically, if we have $k$ entities each with $n$ states, then we have a worst case upper bound of
$(n^{n^{k}})^{k}$ possible candidate models to consider for any abstraction mapping.
For instance, there are $(2^{2^{3}})^{3} = 16777216$ possible Boolean networks consisting of just three entities!
The rest of this section considers techniques for efficiently identifying abstractions and provides
a basis for automating this task.
Initial ideas for implementing these techniques are presented in \cite{Banks09}.

We begin by considering how an abstraction mapping can be applied to an MVN to produce a
set of potential abstraction models.
\\
\\
\addDefn{Definition}{dfn:absMVN}
Let $\absMap = \langle \mapping{\node{1}}, \ldots, \mapping{\node{k}} \rangle$ be an abstraction mapping for
an MVN $\mvn$.
For each entity $\node{i} \in \mvn$ we can abstract the next-state function
$\funcEntity{\node{i}} : \ssEntity{\node{i_1}} \times \dots \times \ssEntity{\node{i_n}} \rightarrow \ssEntity{\node{i}}$
to a (possibly) non-deterministic next-state function
$$\absMap(\funcEntity{\node{i}}) : \mapping{\node{i_1}}(\ssEntity{\node{i_1}}) \times \dots \times \mapping{\node{i_n}}(\ssEntity{\node{i_n}}) \rightarrow \mapping{\node{i}}(\ssEntity{\node{i}})$$
by applying $\absMap$ to its definition in the obvious way.
We say that $\mvn^{A}$ \emph{results from applying} $\absMap$ \emph{to} $\mvn$ iff:
\\
(1) $\mvn^{A}$ has the same entities and neighbourhood structure as $\mvn$;
\\
(2) The state space of each entity $\node{i} \in \mvn^{A}$ is the set $\mapping{\node{i}}(\ssEntity{\node{i}})$;
\\
(3) For each $\node{i} \in \mvn^{A}$ its next-state function
$f^{\mvn^{A}}_{\node{i}} : \mapping{\node{i_1}}(\ssEntity{\node{i_1}}) \times \dots \times \mapping{\node{i_n}}(\ssEntity{\node{i_n}}) \rightarrow \mapping{\node{i}}(\ssEntity{\node{i}})$
is a deterministic restriction of $\absMap(\funcEntity{\node{i}})$.
\\
\\
We define $\absMap(\mvn)$ to be the set of all such MVNs, i.e.
$$
\absMap(\mvn) = \{ \mvn^{A} \ | \ \mvn^{A} \text{ results from applying } \absMap \text{ to } \mvn \}
$$
The trace semantics of $\absMap(\mvn)$ is then defined by
$
\language{\absMap(\mvn)}= \bigcup_{\mvn^{A} \in \absMap(\mvn)} \language{\mvn^{A}}
$
\pfbox
\\

\begin{figure}[h]
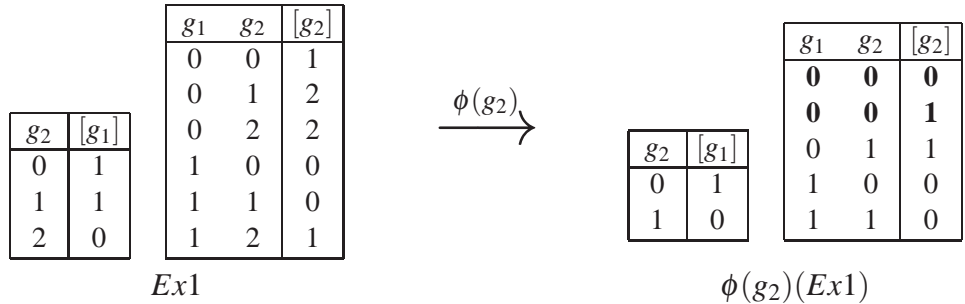

\centering
\begin{tabular}{c@{\qquad}c@{\qquad}c}
    \begin{tabular}{c@{\qquad}c}
      \begin{tabular}[b]{|c|c|}
        \hline $\node{2} $  & $\!\!\nextSt{\node{1} }\!\!$  \\
        \hline
        0 & 1 \\
        1 & 1 \\
        2 & 0 \\
        \hline
      \end{tabular}
        & \hspace{-1.0em}
      \begin{tabular}[b]{|cc|c|}
        \hline $\node{1} $  & $\node{2} $ & $\!\!\nextSt{\node{2} }\!\!$\\
        \hline
        0 & 0 & 1 \\
        0 & 1 & 2 \\
        0 & 2 & 2 \\
        1 & 0 & 0 \\
        1 & 1 & 0 \\
        1 & 2 & 1 \\
        \hline
      \end{tabular}
    \end{tabular}
    &
    \begin{tabular}[b]{c}
        $\mapping{\node{2}}$ \\
        {\Huge $\longrightarrow$}
     \end{tabular}
    &
    \begin{tabular}{c@{\qquad}c}
     \begin{tabular}[b]{|c|c|}
        \hline $\node{2} $  & $\!\!\nextSt{\node{1} }\!\!$  \\
        \hline
        0 & 1 \\
        1 & 0 \\
        \hline
     \end{tabular}
     & \hspace{-1.0em}
     \begin{tabular}[b]{|cc|c|}
        \hline $\node{1} $  & $\node{2} $ & $\!\!\nextSt{\node{2} }\!\!$\\
        \hline
        \textbf{0} & \textbf{0} & \textbf{0} \\
        \textbf{0} & \textbf{0} & \textbf{1} \\
        0 & 1 & 1 \\
        1 & 0 & 0 \\
        1 & 1 & 0 \\
        \hline
    \end{tabular}
   \end{tabular}
    \\
    {\large $\mvnEx1$} & & {\large $\mapping{\node{2}}(\mvnEx1)$}
\end{tabular}
\caption{The (non--deterministic) state transition tables for $\mapping{\node{2}}(\mvnEx1)$ which result from applying
$\mapping{\node{2}}$ to the state transition tables of $\mvnEx1$ (Figure \ref{fig:mvnExample}).
}
\label{fig:mvnExTranslated}
\end{figure}

To illustrate this idea, consider applying the abstraction mapping
$\mapping{\node{2}} =\{0 \mapsto 0, 1 \mapsto 0, 2 \mapsto 1 \}$
to the example MVN $\mvnEx1$ introduced in Section \ref{sec:mvn} (see Figure \ref{fig:mvnExample}).
The resulting  abstracted next-state functions are presented in Figure \ref{fig:mvnExTranslated}. 
The set $\mapping{\node{2}}(\mvnEx1)$ will contain two candidate abstractions in which the
state space for $\node{2}$ is reduced to $\{0, 1 \}$ and whose next-state functions are given by the two possible
interpretations (highlighted in bold) for the abstracted state transition table for $\node{2}$ given in Figure \ref{fig:mvnExTranslated}.

An interesting observation arises by noting that for a given model $\mvn$
and abstraction mapping $\absMap$, the trace semantics of the abstracted MVN
$\language{\absMap(\mvn)}$ is not in general the same as the
abstracted trace semantics $\absMap(\language{\mvn})$.
In fact, it turns out that an important relationship exists between the two, in that
$\language{\absMap(\mvn)}$ will always contain at least
the traces of $\absMap(\language{\mvn})$, as shown by the following theorem.
\\
\\
\addDefn{Theorem}{theorem:subset}
Let $\absMap = \langle \mapping{\node{1}}, \ldots, \mapping{\node{k}} \rangle$
be an abstraction mapping for $\mvn$.
Then we have
$$
\absMap(\language{\mvn}) \subseteq \language{\absMap(\mvn)}.
$$
\textbf{Proof.}
Let $\trace = \left\langle S_{0}, \ldots, S_{n} \right\rangle  \in \language{\mvn}$
be an arbitrary trace, then we need to show that
if $\absMap(\trace)$ is a valid abstracted trace then
$\absMap(\trace) \in \language{\absMap(\mvn)}$.
Let $S_{i} \rightarrow S_{i+1}$ be an arbitrary state step in $\trace$.
Assuming $\mvn$ has $k$ entities then this state step can be broken up into $k$ components
$S_{i} \rightarrow S_{i+1}^{j}$,
for $j = 1, \ldots, k$.
Applying the abstraction mapping to each component gives
$\absMap(S_{i}) \rightarrow \mapping{\node{j}}(S_{i+1}^{j})$.
Clearly, by Definition \ref{dfn:absMVN} there must exist $\mvn^{A} \in \absMap(\mvn)$
whose next-state functions reproduce each of these abstracted component steps
and so is able to reproduce the complete abstracted state step
$\absMap(S_{i}) \rightarrow \absMap(S_{i+1})$.
Since $\absMap(\trace)$ is a valid abstracted trace it follows that we must be able to find
$\mvn^{A} \in \absMap(\mvn)$ which is able to reproduce all the abstracted state steps
$\absMap(S_{i}) \rightarrow \absMap(S_{i+1})$,
for $i=0,\ldots,n-1$.
Thus, we know
$\absMap(\trace) \in \language{\mvn^{A}}$ and so by Definition \ref{dfn:absMVN} we have
$\absMap(\trace) \in \language{\absMap(\mvn)}$
as required.
\pfbox
\\

From this result, it follows that any abstraction of an MVN $\mvn$
must be contained within the set of potential abstractions $\absMap(\mvn)$
as formalized in the corollary below.
\\
\\
\addDefn{Corollary}{cor:subset}
Given two MVNs $\mvni{1}$ and $\mvni{2}$ we have that
$$
\mvni{1} \refines{\absMap} \mvni{2} \ \implies \ \mvni{1} \in \absMap(\mvni{2}).
$$
\textbf{Proof.}
By Definition \ref{def:refinement} we know
$\language{\mvni{1}} \subseteq \absMap(\language{\mvni{2}})$
and so by Theorem \ref{theorem:subset} we have
$\language{\mvni{1}} \subseteq \language{\absMap(\mvni{2})}$.
It therefore follows by Definition \ref{dfn:absMVN} that
$\mvni{1} \in \absMap(\mvni{2})$.
\pfbox
\\

Corollary \ref{cor:subset} provides an important necessary condition for an MVN to be an abstraction of another for a given
abstraction mapping.
It gives us a way of restricting the models that need to be considered when iterating through possible
candidate abstractions for an MVN;
we simply apply the abstraction mapping to the MVN in question and then consider each possible
deterministic model that results from this application. 
This observation results in an exponentially smaller search space and
provides the basis for a more efficient abstraction finding algorithm.

To illustrate the above ideas let us consider finding all the abstractions for $\mvnEx1$ under $\mapping{\node{2}}$,
i.e. calculating the abstraction set $\setAbs{\mvnEx1}{\mapping{\node{2}}}$.
Using the results from Corollary \ref{cor:subset}, we begin by abstracting the state transition tables for $\mvnEx1$
using the given abstraction mapping (shown previously in Figure \ref{fig:mvnExTranslated})
and identifying the potential abstractions contained in $\mapping{\node{2}}(\mvnEx1)$.
We can see that the behaviour of $\node{2}$ is non-deterministic when $\node{1} = 0$ and $\node{2} = 0$.
As such, we have just two possible candidate models $\mvnAI$ and $\mvnAII$ to consider,
shown respectively by Figure \ref{fig:candidatemodels}.(a) and Figure \ref{fig:candidatemodels}.(b)
(where the rules highlighted in bold are the only ones that differ).

\begin{figure}[h]
\vspace*{-0.5em}
\centering
\begin{tabular}{c@{\qquad}c}
    \begin{tabular}{c@{\qquad}c}
     \begin{tabular}[b]{|c|c|}
        \hline $\node{2} $  & $\!\!\nextSt{\node{1} }\!\!$  \\
        \hline
        0 & 1 \\
        1 & 0 \\
        \hline
     \end{tabular}
    & \hspace{-1.0em}
     \begin{tabular}[b]{|cc|c|}
        \hline $\node{1} $  & $\node{2} $ & $\!\!\nextSt{\node{2} }\!\!$\\
        \hline
        \textbf{0} & \textbf{0} & \textbf{0} \\
        0 & 1 & 1 \\
        1 & 0 & 0 \\
        1 & 1 & 0 \\
        \hline
     \end{tabular}
         \vspace*{1.0em}
   \end{tabular}
    & \hspace{1.0em}
   \begin{tabular}{c@{\qquad}c}
    \begin{tabular}[b]{|c|c|}
        \hline $\node{2} $  & $\!\!\nextSt{\node{1} }\!\!$  \\
        \hline
        0 & 1 \\
        1 & 0 \\
        \hline
    \end{tabular}
    & \hspace{-1.0em}
    \begin{tabular}[b]{|cc|c|}
        \hline $\node{1} $  & $\node{2} $ & $\!\!\nextSt{\node{2} }\!\!$\\
        \hline
        \textbf{0} & \textbf{0} & \textbf{1} \\
        0 & 1 & 1 \\
        1 & 0 & 0 \\
        1 & 1 & 0 \\
        \hline
    \end{tabular}
         \vspace*{1.0em}
   \end{tabular}
     \\
   \begin{tabular}{l}
    $\traceState{00} = \left\langle 00, 10, 10  \right\rangle$ \\
    $\traceState{01} = \left\langle 01, 01  \right\rangle$ \\
    $\traceState{10} = \left\langle 10, 10  \right\rangle$ \\
    $\traceState{11} = \left\langle 11, 00, 10, 10  \right\rangle$ \\
  \end{tabular}
   &
  \begin{tabular}{l}
    $\traceState{00} = \left\langle 00, 11, 00  \right\rangle$ \\
    $\traceState{01} = \left\langle 01, 01  \right\rangle$ \\
    $\traceState{10} = \left\langle 10, 10  \right\rangle$ \\
    $\traceState{11} = \left\langle 11, 00, 11  \right\rangle$ \\
  \end{tabular}
  \\
  \\
  (a) Candidate model $\mvnAI$ & (b) Candidate model $\mvnAII$
\end{tabular}

\caption{The state transition tables and trace semantics for candidate models $\mvnAI$ and $\mvnAII$.}
\label{fig:candidatemodels}
\end{figure}
In order to verify whether $\mvnAI$ and $\mvnAII$ are indeed abstractions according to our theory,
we check if their trace semantics are contained within $\mapping{\node{2}}(\language{\mvnEx1})$.
By considering Figure \ref{fig:abstractTrace} and Figure \ref{fig:candidatemodels}
we can observe that $\mvnAI$ is not an abstraction according to Definition \ref{def:refinement},
since $\language{\mvnAI} \not\subseteq \mapping{\node{2}}(\language{\mvnEx1})$;
in other words, its behaviour is not regarded as being consistent with $\mvnEx1$.
On the other hand, we find that $\mvnAII$ is a correct abstraction as
$\language{\mvnAII} \subseteq \mapping{\node{2}}(\language{\mvnEx1})$.
(Indeed, we can see that $\mvnAII$ is precisely the same MVN as $\mvnExB$ which was introduced as an abstraction in the previous section.)
Thus, we have shown that the refinement set \
$\setAbs{\mvnEx1}{\mapping{\node{i}}} \ = \ \{\mvnAII \}$.

It can be observed that exact refinements occur precisely when
the translated MVN has a singleton set of candidate abstraction models,
as shown by the following theorem.
\\
\\
\addDefn{Theorem}{theorem:exact}
Let $\absMap$ be an abstraction mapping for some MVN $\mvn$.
Then we know the following:
\begin{itemize}
  \item [(1)] if $\absMap(\mvn) = \{ \mvn^{A} \}$ is a singleton set, then $\mvn^{A} \exrefines{\absMap} \mvn$;
  \item [(2)] if $\absMap(\mvn)$ is not a singleton set, then no exact abstraction for $\mvn$
  can exist under $\absMap$.
\end{itemize}
\textbf{Proof.}
To prove (1), we observe that if $\absMap(\mvn) = \{ \mvn^{A} \}$ is a singleton set
then for each $\node{i} \in \mvn$ the abstracted next-state function
$\absMap(f_{\node{i}})$ must be deterministic.
This implies that all abstracted traces $\absMap(\sigma)$, for $\sigma \in \language{\mvn}$
must be valid.
Furthermore, by Definition \ref{dfn:absMVN} and Theorem \ref{theorem:subset} it follows that
$\language{\mvn^{A}} = \absMap(\language{\mvn})$ as required.

To prove (2), note that if $\absMap(\mvn)$ contains more than one potential abstraction model
then there must exist at least one abstracted next-state function
$\absMap(f_{\node{i}})$ which is non--deterministic.
This implies there must exist at least one abstracted global state which leads to two or more different traces.
Clearly, either some of these abstracted traces are invalid or $\absMap(\language{\mvn})$ must contain
more traces than any single abstraction model could capture.
Therefore, there cannot exist an exact abstraction for $\mvn$.
\pfbox

\section{Illustrative Biological Examples}
In this section we illustrate the theory and techniques developed in the previous sections by
investigating the existence of abstractions for two published MVN models
for the genetic regulatory network controlling the lysis--lysogeny
switch in the bacteriophage $\lambda$ \cite{Thieffry1995, Chaouiya2008}.
We begin with a brief introduction to the bacteriophage $\lambda$ (see \cite{opp05} for a more detailed introduction).

The temperate bacteriophage $\lambda$ is a
virus which infects the bacteria \emph{Escherichia coli} \cite{Thomas1990, opp05}.
After infection of the host cell, a decision is made by $\lambda$ based
on environmental factors between two very different methods of
reproduction, namely the \emph{lytic} and \emph{lysogenic} cycles \cite{Thomas1990}.
In most cases, $\lambda$ enters the \emph{lytic cycle},
where it generates as many new viral particles as the host cell
resources allow before producing an enzyme to lyse the cell wall,
releasing the new phage into the environment.
Alternatively, the $\lambda$ DNA may integrate into the host DNA and enter the \emph{lysogenic cycle}.
Importantly, genes expressed in the $\lambda$ DNA synthesize a
repressor which blocks expression of other phage genes including
those involved in its own excision.
As such, the host cell establishes an immunity to external infection from other
phages, and the phage $\lambda$ is able to lie dormant, replicating with
each subsequent cell division of the host.

\subsection{The Two Entity Core Regulatory Model}

\begin{figure}[h]
\centering
\begin{tabular}{c@{\qquad}c}
   \includegraphics[width=0.4\textwidth,keepaspectratio]{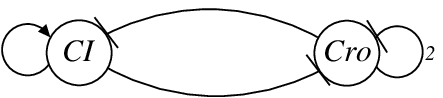}
        &
   \begin{tabular}[b]{l l}
     $\traceState{00} = \left\langle 00, 11, 00  \right\rangle$ &
     $\traceState{10} = \left\langle 10, 10  \right\rangle$ \\
     $\traceState{01} = \left\langle 01, 02, 01  \right\rangle$ &
     $\traceState{11} = \left\langle 11, 00, 11  \right\rangle$ \\
     $\traceState{02} = \left\langle 02, 01, 02  \right\rangle$ &
     $\traceState{12} = \left\langle 12, 01, 02, 01  \right\rangle$ \\

   \end{tabular}
   \\
   \\
    (a) Network structure & (c) Trace semantics
   \\
   \\
   \begin{tabular}[b]{c@{\qquad}c@{\qquad}c}
      \begin{tabular}[b]{|cc|c|}
        \hline $\CI$  & $\Cro$ & $\!\!\nextSt{\CI}\!\!$\\
        \hline
        0 & 0 & 1 \\
        0 & 1 & 0 \\
        0 & 2 & 0 \\
        1 & 0 & 1 \\
        1 & 1 & 0 \\
        1 & 2 & 0 \\
        \hline
     \end{tabular}
    & \hspace{-1.0em}
     \begin{tabular}[b]{|cc|c|}
        \hline $\CI$  & $\Cro$ & $\!\!\nextSt{\Cro}\!\!$\\
        \hline
        0 & 0 & 1 \\
        0 & 1 & 2 \\
        0 & 2 & 1 \\
        1 & 0 & 0 \\
        1 & 1 & 0 \\
        1 & 2 & 1 \\
        \hline
     \end{tabular}
   \end{tabular}
  &
   \begin{tabular}[b]{c}
      \\
      \includegraphics[width=0.27\textwidth,keepaspectratio]{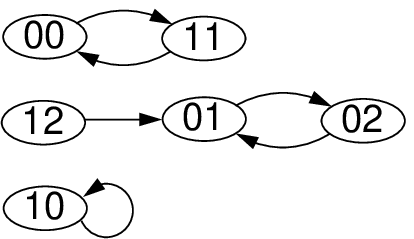}
      \\
      \\
   \end{tabular}
  \\
  (b) State transition tables & (d) Graphical representation of traces\\
\end{tabular}
\caption{
Formal definition and trace semantics for the MVN model $\PLI$ of the core regulatory mechanism
for the lysis-lysogeny switch in bacteriophage $\lambda$ (taken from \cite{Thieffry1995}).}
\label{fig:exPhage1}
\end{figure}

A simple MVN model of the core regulatory mechanism for the
lysis--lysogeny switch was proposed in \cite{Thieffry1995}.
This model, which we denote as $\PLI$, is presented in Figure \ref{fig:exPhage1}
and is based on the cross--regulation between two regulatory genes, $\CI$ (the repressor gene) and $\Cro$.
It can be seen that $\Cro$ inhibits the expression of $\CI$ and
at higher levels of expression, also inhibits itself.
The gene $\CI$ inhibits the expression of $\Cro$ while
promoting its own expression.
The full synchronous trace semantics $\language{\PLI}$ for this MVN is presented in Figure \ref{fig:exPhage1}.(c).
We can see from the state transition graph in Figure \ref{fig:exPhage1}.(d) that $\PLI$ has three attractor cycles,
where the attractor cycle $10 \updateStep 10$ represents the lysogenic cycle since the repressor gene $\CI$ is fully expressed
and $01\updateStep 02 \updateStep 01$ represents the lytic cycle.

In order to identify an abstraction for $\PLI$ we begin by selecting an appropriate
state mapping $\mapping{\Cro} : \{0,1,2\} \rightarrow \{0, 1\}$
for the only non-Boolean entity $\Cro$.
We use our understanding of the behaviour of $\Cro$ to define the following state mapping
$$
\mapping{\Cro} = \{0 \mapsto 0, 1 \mapsto 1, 2 \mapsto 1\}.
$$
We can then view $\mapping{\Cro}$ as an abstraction mapping and
following the approach in Section \ref{sec:ident}, we restrict the abstraction search space
by applying the abstraction mapping $\mapping{\Cro}$ to $\PLI$.
This results in a set $\mapping{\Cro}(\PLI)$ which
contains two candidate abstraction models.
\begin{figure}[h]
\centering
\begin{tabular}[b]{c@{\qquad}c}
   \begin{tabular}[b]{c@{\qquad}c@{\qquad}c}
      \begin{tabular}[b]{|cc|c|}
        \hline $\CI$  & $\Cro$ & $\!\!\nextSt{\CI}\!\!$\\
        \hline
        0 & 0 & 1 \\
        0 & 1 & 0 \\
        1 & 0 & 1 \\
        1 & 1 & 0 \\
        \hline
     \end{tabular}
    & \hspace{-1.0em}
     \begin{tabular}[b]{|cc|c|}
        \hline $\CI$  & $\Cro$ & $\!\!\nextSt{\Cro}\!\!$\\
        \hline
        0 & 0 & 1 \\
        0 & 1 & 1 \\
        1 & 0 & 0 \\
        1 & 1 & 0 \\
        \hline
     \end{tabular}
   \end{tabular}
   &
   \begin{tabular}[b]{l}
      $\traceState{00} = \left\langle 00, 11, 00  \right\rangle$ \\
      $\traceState{01} = \left\langle 01, 01  \right\rangle$ \\
      $\traceState{10} = \left\langle 10, 10  \right\rangle$ \\
      $\traceState{11} = \left\langle 11, 00, 11  \right\rangle$ \\
      \\
   \end{tabular}
\end{tabular}
\caption{
Abstraction model $\PLIab$ for $\PLI$ and associated trace semantics $\language{\PLIab}$.}
\label{fig:abstractPhage1}
\end{figure}
It turns out that only one of these is a correct abstraction and
we present this abstraction model $\PLIab$ in Figure \ref{fig:abstractPhage1}.
It is straightforward to check that the trace semantics of $\PLIab$
(see Figure \ref{fig:abstractPhage1})
is indeed consistent with the abstracted trace semantics of $\PLI$
(see Figure \ref{fig:transTracesPhage1}),
i.e. $\language{\PLIab} \subseteq \mapping{\Cro}(\language{\PLI})$.
Thus, we know $\PLIab \refines{\mapping{\Cro}} \PLI$ holds.
\begin{figure}[h]
\centering
\begin{tabular}[b]{l l}
$\mapping{\Cro}(\traceState{00}) = \left\langle 00, 11, 00  \right\rangle$ &
$\mapping{\Cro}(\traceState{10}) = \left\langle 10, 10  \right\rangle$ \\
$\mapping{\Cro}(\traceState{01}) = \left\langle 01, 01  \right\rangle$ &
$\mapping{\Cro}(\traceState{11}) = \left\langle 11, 00, 11  \right\rangle$ \\
$\mapping{\Cro}(\traceState{02}) = \left\langle 01, 01  \right\rangle$ &
$\mapping{\Cro}(\traceState{12}) = \left\langle 11, 01, 01  \right\rangle$ \\
\end{tabular}
\caption{
The traces $\mapping{\Cro}(\language{\PLI})$ resulting from abstracting the traces of $\PLI$
using $\mapping{\Cro}$.}
\label{fig:transTracesPhage1}
\end{figure}

It can be seen that the abstraction $\PLIab$ acts as a good approximation to the behaviour of the original
MVN $\PLI$ and in particular, we can see that the abstraction has captured all three attractor cycles that
were present in $\PLI$.

\subsection{The Four Entity Regulatory Model}

\begin{figure}[h]
\centering
\begin{tabular}{c@{\qquad}c@{\qquad}c}
   \includegraphics[width=0.35\textwidth,keepaspectratio]{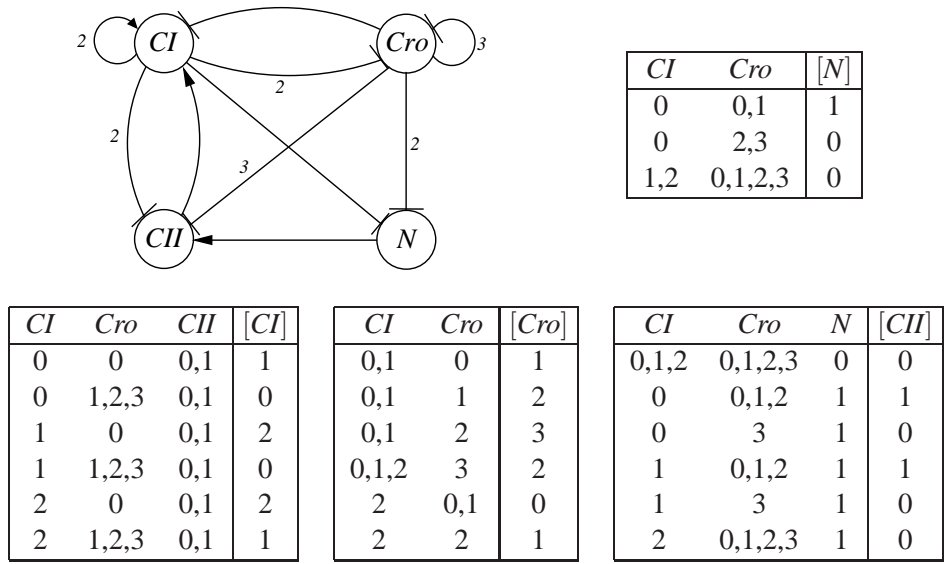}
        & &
   \begin{tabular}[b]{c}
    \begin{tabular}[b]{|cc|c|}
        \hline $\CI$  & $\Cro$ & $\!\!\nextSt{\N}\!\!$\\
        \hline
        0   & 0,1     & 1 \\
        0   & 2,3     & 0 \\
        1,2 & 0,1,2,3 & 0 \\
        \hline
     \end{tabular}
     \\
     \\
     \\
    \end{tabular}
\end{tabular}
\\
\vspace{3mm}
\begin{tabular}{c@{\qquad}c@{\qquad}c}
     \begin{tabular}[b]{|ccc|c|}
        \hline $\CI$  & $\Cro$ & $\CII$ & $\!\!\nextSt{\CI}\!\!$\\
        \hline
        0 & 0     & 0,1 & 1 \\
        0 & 1,2,3 & 0,1 & 0 \\
        1 & 0     & 0,1 & 2 \\
        1 & 1,2,3 & 0,1 & 0 \\
        2 & 0     & 0,1 & 2 \\
        2 & 1,2,3 & 0,1 & 1 \\
        \hline
     \end{tabular}
     & \hspace{-1.0em}
     \begin{tabular}[b]{|cc|c|}
        \hline $\CI$  & $\Cro$ & $\!\!\nextSt{\Cro}\!\!$\\
        \hline
        0,1   & 0   & 1 \\
        0,1   & 1   & 2 \\
        0,1   & 2   & 3 \\
        0,1,2 & 3   & 2 \\
        2     & 0,1 & 0 \\
        2     & 2   & 1 \\
        \hline
     \end{tabular}
     & \hspace{-1.0em}
     \begin{tabular}[b]{|ccc|c|}
        \hline $\CI$  & $\Cro$ & $\N$ & $\!\!\nextSt{\CII}\!\!$\\
        \hline
        0,1,2 & 0,1,2,3 & 0 & 0 \\
        0     & 0,1,2   & 1 & 1 \\
        0     & 3       & 1 & 0 \\
        1     & 0,1,2   & 1 & 1 \\
        1     & 3       & 1 & 0 \\
        2     & 0,1,2,3 & 1 & 0 \\
        \hline
     \end{tabular}
\end{tabular}
\caption{
An extended MVN model $\PLII$ of the control mechanism for the lysis-lysogeny switch
in bacteriophage $\lambda$ (taken from \cite{Thieffry1995}).}
\label{fig:exPhage2}
\end{figure}

The core regulatory model presented above was extended in \cite{Thieffry1995} to
take account of the actions of two further regulatory genes, $\CII$ and $\N$.
The resulting four entity MVN model $\PLII$ is presented in Figure \ref{fig:exPhage2}
(note that the state transition tables presented use a shorthand notation where an entity is allowed to
be in any of the states listed for it in a particular row).
This MVN is more detailed than $\PLI$  and contains two entities with non-Boolean state spaces,
namely $\CI$ with states $\{0,\ldots,2\}$ and
$\Cro$ with states $\{0,\ldots,3\}$.
The resulting state space for the model consists of $48$ global states and for this reason
we do not reproduce its trace semantics here.
Instead, we simply note that $\PLII$ has the following three attractor cycles
(where the first corresponds to the lytic cycle and the remaining two to the lysogenic cycle)
$$
0300 \updateStep 0200 \updateStep 0300  , \ \ \ \ \ \ 1000 \updateStep 2100 \updateStep 1000,
\ \ \ \ \ \ 2000 \updateStep 2000
$$

We begin by looking to abstract the non-Boolean entities $\CI$ and $\Cro$ by defining appropriate state mappings.
After considering the model, we define
the following state mappings
$$
\mapping{\CI} = \{0 \mapsto 0, 1 \mapsto 1, 2 \mapsto 1\}, \ \ \ \
\mapping{\Cro} = \{0 \mapsto 0, 1 \mapsto 1, 2 \mapsto 1, 3 \mapsto 1\}.
$$
which we use to define the abstraction mapping
$\absMap = \langle \mapping{\CI}, \mapping{\Cro}, I_{\CII}, I_{\N} \rangle$.
Again, following the approach presented in Section \ref{sec:ident} we first apply this abstraction
mapping to $\PLII$ resulting in the set $\absMap(\PLII)$ of candidate abstraction models.
By analysing $\absMap(\PLII)$ we are able to establish that there are
256 possible candidate abstraction models (we have 4 choices for $\CI$, 4 choices for $\Cro$,
8 choices for $\CII$,
and 2 choices for $\N$).
After investigating these candidate models we were able to identify two abstractions for $\PLII$
under $\absMap$, denoted $\PLIIabI \refines{\absMap} \PLII$ and
$\PLIIabII \refines{\absMap} \PLII$, which are presented in Figure \ref{fig:exPhage2Abs}.
Interestingly, both abstractions appear to capture the key behaviour of $\PLII$ in the sense that
both contain the attractor cycles
$0100 \updateStep 0100$ and $1000 \updateStep 1000$
which correspond to those present in $\PLII$.
\begin{figure}[h]
\centering
\begin{tabular}{c@{\qquad}c@{\qquad}c}
   \begin{tabular}[b]{|ccc|c|}
        \hline $\CI$  & $\Cro$ & $\CII$ & $\!\!\nextSt{\CI}\!\!$\\
        \hline
        0 & 0 & 0,1 & 1 \\
        0 & 1 & 0,1 & 0 \\
        1 & 0 & 0,1 & 1 \\
        1 & 1 & 0,1 & 0 \\
        \hline
   \end{tabular}

   &

   \begin{tabular}[b]{c}
     \begin{tabular}[b]{|cc|c|}
        \hline $\CI$  & $\Cro$ & $\!\!\nextSt{\N}\!\!$\\
        \hline
        0   & 0   & 1 \\
        0   & 1   & 0 \\
        1   & 0,1 & 0 \\
        \hline
     \end{tabular}
   \\
   \\
     \begin{tabular}[b]{|cc|c|}
        \hline $\CI$  & $\Cro$ & $\!\!\nextSt{\Cro}\!\!$\\
        \hline
        0   & 0,1  & 1 \\
        1   & 0    & 0 \\
        1   & 1    & 1 \\
        \hline
     \end{tabular}
   \end{tabular}

   &

     \begin{tabular}[b]{|ccc|c|}
        \hline $\CI$  & $\Cro$ & $\N$ & $\!\!\nextSt{\CII}\!\!$\\
        \hline
        0,1 & 0,1 & 0 & 0 \\
        0   & 0   & 1 & 1 \\
        \textbf{0}    & \textbf{1}   & \textbf{1} & \textbf{1} \ or \ \textbf{0} \\
        1   & 0 & 1 & 0 \\
        1   & 1 & 1 & 1 \\
        \hline
     \end{tabular}
\end{tabular}
\caption{
The transition tables for the two abstractions $\PLIIabI$ and $\PLIIabII$ identified for $\PLII$
under $\absMap$, where all the transition tables are the same except for $\CII$
where  $011 \updateStep 1$ for abstraction $\PLIIabI$ but  $011 \updateStep 0$ for abstraction $\PLIIabII$.}
\label{fig:exPhage2Abs}
\end{figure}

\section{Conclusions}
In this paper we have developed an abstraction theory for MVN models based
on the idea of using an abstraction mapping to relate the reduced state space of
an abstraction to the original model.
The problem of identifying suitable abstractions for an MVN was discussed and
some initial ideas for restricting the number of candidate abstraction models that need to be considered were proposed.
We showed that abstractions can be used to analyse an MVN since they preserve reachability properties and importantly, since all
the attractor cycles of an abstraction will correspond to attractor cycles in the original model.
This work was motivated by the need to be able to relate MVN models at different levels of
abstraction and in particular, the idea of abstracting an MVN to a simpler model which
is more amenable to analysis and visualization techniques.
The abstraction theory presented can also be seen as providing a framework for an incremental
refinement approach to constructing MVNs.

We illustrated the abstraction theory and techniques developed by considering two
examples based on published MVN models of the genetic regulatory
network for the lysis-lysogeny switch in phage $\lambda$~\cite{Thieffry1995, Chaouiya2008}.
We considered a simple two entity model and then an extended model that contained four entities (two of which
were non-Boolean).
In both cases we were able to identify meaningful Boolean abstractions which
captured the key attractor cycles contained in the original models.

Further work is now needed to build on the ideas presented in Section \ref{sec:ident}
to develop tool support for automatically checking and identifying abstractions.
Initial ideas for such tool support have been presented in \cite{Banks09} and work is
on going to develop efficient algorithmic solutions to support the abstraction process.
Other researchers have considered abstracting MVNs by reducing the number of regulatory
entities while preserving important model dynamics (see for example \cite{Naldi09,Velizcuba2009}).
It would be interesting to consider combining such an approach with the abstraction theory we have developed here.
Finally, we note that extending the abstraction theory to asynchronous MVN models is an interesting but challenging area of future work.
In particular, ways of coping with the
non-deterministic choices inherent in the dynamic behaviour of asynchronous models will be needed.

\medskip
\noindent\textbf{Acknowledgments.} We would like
to thank the \textsc{Epsrc} for supporting R. Banks during part of this work.
We are also very grateful to Maciej Koutny, Hanna Klaudel, and Michael Harrison for their help
and advice during the preparation of this paper.
Finally, we would like to thank the anonymous referees for their helpful comments.

\bibliographystyle{eptcs}

\begin{thebibliography}{99}

\bibitem{amk99} T. Akutsu, S. Miyano and S. Kuhara,
    Identification of genetic networks from small number
    of gene expression patterns under the Boolean network
    model, \emph{Proc. of Pac. Symp. on Biocomp.}, 4:17--28, 1999.
%
\bibitem{Banks09}
R. Banks. \emph{Qualitatively Modelling Genetic Regulatory Networks: Petri Net Techniques and Tools}. 
Ph. D. Dissertation, School of Computing Science, University of Newcastle upon Tyne, 2009.
%
\bibitem{Bensalem1998}
S. Bensalem, Y. Lakhnech, and S. Owre.
Computing Abstractions of Infinite State Systems Compositionally and Automatically.
In: \emph{Proc. of the 10th Int. Conference on Computer Aided Verification},
Lecture Notes In Computer Science 1427, pages 319--331, Springer-Verlag, 1998.
%
\bibitem{Bower01}
J. Bower, and H. Bolouri.
\emph{Computational Modelling of Genetic and Biochemical Networks}, MIT Press, 2001.
%
\bibitem{Chaouiya2008}
C. Chaouiya, E. Remy, and D. Thieffry.
Petri Net Modelling of Biological Regulatory Networks.
\emph{Journal of Discrete Algorithms}, 6(2):165--177, 2008.
%
\bibitem{Clarke1994}
E. M. Clarke, O. Grumberg, and D. E. Long.
Model Checking and Abstractions.
\emph{ACM Transactions on Programming Languages and Systems}, 16(5):1512 - 1542, 1994.
%
\bibitem{deJong02}
H. de Jong. Modeling and simulation of genetic regulatory systems: a literature review.
\emph{Journal of Computational Biology}, 9:67--103, 2002.
%
\bibitem{Drossel2005}
B. Drossel, T. Mihaljev, and F. Greil.
Number and length of attractors in a critical Kauffman model with connectivity one.
\emph{Physical Review Letters}, 94(8), 2005.
%
\bibitem{Harvey1997}
I. Harvey and T. Bossomaier.
Time Out of Joint: Attractors in Asynchronous Random Boolean Networks.
In: P. Husbands and I. Harvey (eds.), \emph{Proc. of ECAL97}, pages 67--75, MIT Press 1997.
%
\bibitem{huang2000}
S. Huang and D. Ingber.
Shape-dependent control of cell growth, differentiation, and
apoptosis: Switching between attractors in cell regulatory networks.
\emph{Experimental Cell Research}, 261(1):91--103, 2000.
%
\bibitem{kauffman1993} S. Kauffman.
\emph{The origins of order: Self-organization and selection in evolution.}
Oxford University Press, New York, January 1993.
%
\bibitem{Mishchenko2002}
A. Mishchenko and R. Brayton.
Simplification of non-deterministic multi-valued networks.
In: \emph{ICCAD '02: Proc. of the 2002 IEEE/ACM Int.  Conference on Computer-aided design},
pages 557--562, 2002.
%
\bibitem{Naldi09}
A. Naldi, E. Remy, D. Thieffry, and C. Chaouiya.
A Reduction of Logical Regulatory Graphs Preserving Essential Dynamical Properties.
In: \emph{Proc. of CMSB '09}, Lecture Notes in Bioinformatics 5688, pages 266 - 280,
Springer-Verlag, 2009.
%
\bibitem{opp05} A.B. Oppenheim, O. Kobiler, J. Stavans, D. L. Court, and S. L. Adhya.
Switches in bacteriophage $\lambda$ development.
\emph{Annual Review of Genetics}, 39:4470--4475, 2005.
%
\bibitem{Rudell1987}
R. Rudell and A. Sangiovanni-Vincentelli.
Multiple-Valued Minimization for PLA Optimization.
\emph{IEEE Transactions on Computer-Aided Design}, CAD-6, 1987.
%
\bibitem{Schaub2007}
M. Schaub, T. Henzinger, and J. Fisher.
Qualitative networks: A symbolic approach to analyze bio-logical signaling networks.
\emph{BMC Systems Biology}, 1:4, 2007.
%
%
\bibitem{Thieffry1995}
D. Thieffry and R. Thomas.
Dynamical behaviour of biological regulatory networks - II.
Immunity control in bacteriophage lambda.
\emph{Bulletin of Mathematical Biology}, 57:277--295, 1995.
%
\bibitem{Thomas1990}
R. Thomas and R. D'Ari. \emph{Biological Feedback}, CRC Press, 1990.
%
\bibitem{Thomas1995}
R. Thomas, D. Thieffry and M. Kaufman.
Dynamical Behaviour of Biological Regulatory Networks - I. Biological Role of Feedback Loops and Practical use of the Concept of Loop-Characteristic State.
\emph{Bulletin of Mathematical Biology}, 57:247--276, 1995.
%
\bibitem{Velizcuba2009}
A. Veliz--Cuba. Reduction of Boolean Networks. 
http://arxiv.org/abs/0907.0285, submitted 2009. 
\end{thebibliography}

\end{document}